\documentclass{tufte-handout}

\usepackage{enumitem}
\usepackage{amsmath}

\usepackage{graphicx}
\setkeys{Gin}{width=\linewidth,totalheight=\textheight,keepaspectratio}
\graphicspath{{graphics/}}

\usepackage{booktabs}

\usepackage{units}

\usepackage{fancyvrb}
\fvset{fontsize=\normalsize}

\usepackage{multicol}

\usepackage{lipsum}

\usepackage{hyperref} 
\usepackage{enumitem}
\setlist{nosep}

\title{Visualizing Music Genres using a Topic Model}
\author{Swaroop Panda, Vinay P. Namboodiri, Shatarupa Thakurta Roy}
\date{Indian Institute of Technology Kanpur}  
\begin{document}

\maketitle

\begin{abstract}
Music Genres serve as an important meta-data in the field of music information retrieval and have been widely used for music classification and analysis tasks. Visualizing these music genres can thus be helpful for music exploration, archival and recommendation. Probabilistic topic models have been very successful in modelling text documents. In this work, we visualize music genres using a probabilistic topic model. Unlike text documents, audio is continuous and needs to be sliced into smaller segments. We use simple MFCC features of these segments as musical words. We apply the topic model on the corpus and subsequently use the genre annotations of the data to interpret and visualize the latent space.
\end{abstract}

\section{Introduction}
It is suggested by \cite{wells1992origins} that musicologists identify music genres as pieces of music that have a similar musical language or structure. Music genre analysis tasks are widely pursued in the music information retrieval community. However, genre visualizations have not caught enough attention. Music Genre Visualization can be very helpful in modern technologies such as custom music playlists and recommendation systems. Probabilistic Topic Models \cite{blei2012probabilistic}, have found wide applications in the field of Natural Language processing. We use an unsupervised topic model on music genres data for visualization. For our work we use raw music files, in .wav format. Unlike text documents, raw music data has no discrete components such as words. To create a text-like corpus, we slice the audio data into smaller segments. We use MFCC features of these smaller slices as the representation. Further, to build a corpus, we create a feature dictionary by using the k-means algorithm. Also, in text documents, the inferred topics form a collection of words and hence are straightforward to interpret. In our case, musical words, which are mere MFCC feature arrays, lack inherent meaning and cannot be interpreted. We interpret the latent space of the topic model using genre annotations in the dataset.

\section{Related Work}
\label{sec:page_size}
The work in \cite{cooper2006visualization} presented a summary of and had discussed some audio visualization techniques in MIR which are mostly signal processing based. Topic Models have also been applied on audio in \cite{kim2012latent} for audio classification and \cite{kim2009acoustic} for audio information retrieval. The work in \cite{shalit2013modeling} uses a topic model to model musical influence. The work in \cite{huprobabilistic} applies a topic model on musical notes for key analysis.The work in \cite{hirai2016musicmixer} and \cite{hirai2018latent} also use this models on music in the context of DJ performance. Similarly, the work in \cite{hariri2012context} uses topic models on music tags for use in recommendation systems.
We use the fault-filtered GTZAN \cite{tzanetakis2002musical} dataset for genre analysis which is popular in MIR community. 

\section{The Probabilistic Topic Model}
\label{sec:pagestyle}

Probabilistic topic models are based upon an idea that documents are mixture of topics and topics are probability distributions over words. These words come from a fixed size vocabulary. To make a new document, one chooses a distribution of topics, then chooses a topic from this distribution and finally draws a word from the chosen topic. The Latent Dirichlet Allocation (LDA) inverts this generative process and thus infers the set of topics that were that useful in generating the document. The plate notation also helps understand the generative process.

Using some mathematical notations, the generative process can be defined as follows. Let $K$ be the specified number of topics, $V$ the size of the vocabulary, $\alpha$ a positive $K$-vector and $\eta$ a scalar. Let ${Dir_V}$($\alpha$) denote a $V$-dimensional dirichlet with a vector parameter $\vec{\alpha}$ and Let ${Dir_K}$($\eta$) denote a K-dimensional dirichlet with a scalar parameter $\eta$.

\begin{enumerate}
\item For each topic
    \begin{itemize}
        \item Draw a distribution over words $\beta_K$ over ${Dir_V}$($\eta$)  
    \end{itemize}
    
\item For each document 
    \begin{itemize}
        \item Draw a vector of topic proportions $\theta_d$ over $Dir{(\alpha)}$
        \item For each word
        \begin{itemize}
            \item Draw a topic assignment ${Z_{d,n}}$ over \\Mult(${\theta_d}$),${Z_{d,n}}$ $\epsilon$ ${1,2,...,K}$  
            \item Draw a word ${W_{d,n}}$ over\\ Mult($\beta_{Z_{d,n}}$), ${W_{d,n}}$ $\epsilon$ ${1,2,...,V}$
        \end{itemize}
    \end{itemize} 
\end{enumerate}

\begin{figure}[h]
 \centerline{\framebox{
 \includegraphics[width=7cm]{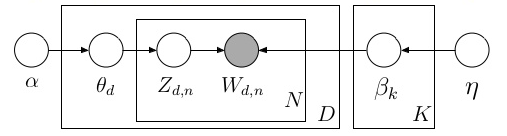}}}
 \caption{Plate Notation for a Topic Model}
 \label{fig:tm8}
\end{figure}

The above figure describes the generative process of the LDA algorithm using a what is called as a plate notation. The shaded variable ${W_{d,n}}$ is observable while all other variables are latent. The direction of the arrows help in understanding the generative process of the algorithm.

\section{Applying Topic Models on Music}
\label{sec:pagestyle01}

The main challenge with the application of topic models in the music (with raw audio files) is to represent the audio in a text-document like corpus. The intent of the work is to interpret the latent space of the topic model using music genres. This interpretation would help giving a probabilistic genre annotation to a song. For example a song may belong 60 \% to Blues, 15 \% to Jazz and 25 \% to Pop genres. To enable such a probabilistic assignment, we build basic genre buckets consisting of at least 3 genres. We do this since a mixture containing all the 10 genres would be very large and obfuscating for the listener to meaningfully interpret. The rationale used to bucket the genres is roughly based on the histories and the musical form of these genres. The first bucket consists of Rock, Metal and Pop genres; the second of Blues, Jazz and Country genres and the final bucket consists of Reggae, Disco and Hip-Hop genres. The songs were clipped down to 0.10 seconds clips. The MFCC features of these clips were then calculated. We then use a K means clustering algorithm on the MFCC features to build the dictionary. It partitions the data into k clusters, where each data point belongs to a cluster and the cluster mean serves as its prototype. The librosa package \cite{mcfee2015librosa} was used for the computation and the pre-processing while the models are implemented using gensim package. An algorithmic description of the process is given as below.

\begin{enumerate}
\item For each song
    \begin{itemize}
        \item Discretize into clips of 0.10 secs 
        \item Calculate MFCC features of these clips.
    \end{itemize}
    
\item For all songs in a genre bucket 
    \begin{itemize}
        \item Calculate K-means (K=3) of the MFCC feature vectors 
        \item Map each MFCC feature vector (and thus the song) to any of the K-clusters
        \item Apply the textual probabilistic topic model
    \end{itemize} 
\end{enumerate}

\section{Interpretation of the Latent Space}

Unlike text documents, where topics are interpreted as a mixture of words; the acoustic topic model has topics which are mixtures of cluster means. These cluster means are prototypes of the nearest datapoints(the audiofiles) and thus lack meaning. It is thus essential to assign a suitable meaning to these cluster means. The first part of the interpretation involves understanding the cluster means in terms of music genre. The cluster means are constructed from the MFCC arrays. The cluster means hence can be mapped to and from these audio files and linked with the genre annotations. For instance, lets say that 3 audio files, audio1, audio4 and audio7 make up a cluster mean. We get back to the dataset and find out that audio1 belongs to the Blues genre, audio4 belongs to the Country genre and the audio7 belongs to the Blues genre. Hence, the genres associated with cluster means becomes {Blues, Country,Blues}. In math, the cluster centers (or terms) can be described as the following,

\begin{equation}
\begin{aligned}
    cluster mean1 & = \{Blues,Country,Blues\} \\
          & = 0.67Blues + 0.33 Country\\
    cluster mean2 & = \{Blues, Jazz, Jazz, Country\}\\
          & = 0.25Blues + 0.5Jazz + 0.25Country\\
          \end{aligned}
\end{equation}



Once we interpret cluster means in terms of music genres, we can conveniently represent the topics in terms of music genres. The topic space consists of cluster means and an associated probability value. The cluster means can now be defined as music genres with their proportions.

\begin{equation}
\begin{split}
    Topic1&= prob1*cluster mean1 + prob2*cluster mean2 \\
           &= (prob1*0.67+prob2*0.25)Blues +\\&(prob1*0.33+prob2*0.25)Country\\&+ (prob2*0.5)Jazz \\
\end{split}
\end{equation}

\subsection{Document-Topic Proportions}

Once the topic space has been interpreted, the document-topic proportions can also be made sense of. The document topic proportions from the topic model are probability values of the inferred topics present in each document. In this context, the document topic proportions can provide with the proportions of different music genres present within the musical document, that is, a song. 

\begin{equation}
\begin{split}
    Doc1&= prob1*topic1 + prob2*topic2 \\
           &= prob1*(0.33Blues + 0.33Country+ 0.34Jazz) \\&+ prob2*(0.5Jazz + 0.5Country)\\
          &= (prob1*0.33)Blues + (prob1*0.34+prob2*0.5)Jazz \\&+ (prob1*0.33+prob2*0.5)Country\\
\end{split}
\end{equation}


\subsection{The Term-Topic Proportions}

The term-topic proportions suggest the topic distributions in each term. So the term-topic distribution represents the genres present in each term, or cluster center. This is represented in an equation as,

\begin{equation}
\begin{split}
    Term1&= prob1*topic1 + prob2*topic4 \\
          &= prob1*(0.33Blues + 0.50Country+ 0.17Jazz) \\&+ prob2*(0.5Blues + 0.5Jazz)\\
          &= (prob1*0.33+prob2*0.5)Blues + (prob1*0.17+prob2*0.5)Jazz \\&+ (prob2*0.5)Country\\
\end{split}
\end{equation}

The term-topic proportions is an interesting insight derived from the topic models. The terms are cluster centers derived from MFCC feature arrays. These feature arrays are in turn derived from small segments of a song. So the term-topic proportion suggest the genre proportions of each small segment of a song. For instance, 3 small segments of a song can belong to Blues-Classical, Jazz-Pop-Classical or Disco-Rock genres in some proportions.

\section{Evaluating the Topic Model}
We evaluate the model using a genre classification task. We use the model to get the document-topic proportions of every document(song). We use these document-topic proportions as a representation for each song. We use genre labels from the fault-filtered GTZAN dataset, divide the data into train-test sets and perform the classification task using a SVM. We also test our model with different number of topics to look for the optimal number of topics that best capture the genre bucket.

\begin{table}[h!]
\begin{center}
\begin{tabular}{ | l | l | l | l | p{0.6cm} |}
\hline
     &2& 3 &4&5  \\ \hline
    1 &0.47&0.53 &0.58 & 0.53 \\ \hline
    2 &0.36 &0.38 &0.35&0.40    \\ \hline
    3 & 0.53&0.46 &0.48 & 0.50   \\ \hline

\hline
\end{tabular}
 \caption{Accuracies obtained for different number of topic terms. The rows represent the genre bucket, while the columns represent the number of topics}
\end{center}
 \end{table}

\section{Music Genre Visualization}
Using the topic model, we can get a probabilistic genre labels of different songs(from document-topic proportions) along with progressive genre visualizations(from term-topic proportions).







\begin{figure}[!h]
	\centering
		\centering
		\includegraphics[width = 8cm]{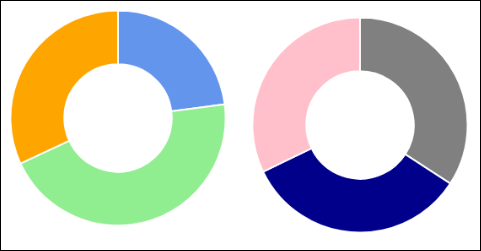}
		\caption{The above is a visualization of two topics emerging out of genre buckets 1 and 2 respectively. The colours represent the music genres and the doughnut partitions show genre proportions within a topic.}
		\end{figure}
	\begin{figure}[!h]
		\centering
		\includegraphics[width = 9cm]{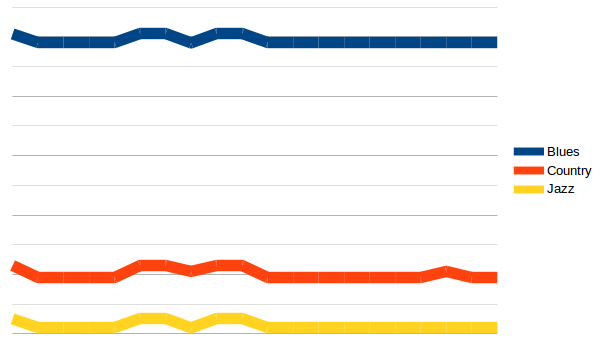}
		\caption{The progressive viz of a song segment. The x-axis denotes time and the y-axis the genre proportions. The above figure displays the changing genre proportions as the song progresses. }
	\end{figure}

\section{Conclusion}
In this work, we applied a Probabilistic Topic Model on raw Music data and used the available genre annotations to interpret the latent space. We used the inferred parameters of the Topic Model to visualize the Music Data and also proposed some genre visualizations. Further, it would be interesting to explore other audio features and genre data for visualization using the Topic Model.

\bibliography{sample-handout}

\begin{thebibliography}{12}
\providecommand{\natexlab}[1]{#1}
\providecommand{\url}[1]{\texttt{#1}}
\expandafter\ifx\csname urlstyle\endcsname\relax
  \providecommand{\doi}[1]{doi: #1}\else
  \providecommand{\doi}{doi: \begingroup \urlstyle{rm}\Url}\fi

\bibitem[Blei(2012)]{blei2012probabilistic}
David~M Blei.
\newblock Probabilistic topic models.
\newblock \emph{Communications of the ACM}, 55\penalty0 (4):\penalty0 77--84,
  2012.

\bibitem[Cooper et~al.(2006)Cooper, Foote, Pampalk, and
  Tzanetakis]{cooper2006visualization}
Matthew Cooper, Jonathan Foote, Elias Pampalk, and George Tzanetakis.
\newblock Visualization in audio-based music information retrieval.
\newblock \emph{Computer Music Journal}, 30\penalty0 (2):\penalty0 42--62,
  2006.

\bibitem[Hariri et~al.(2012)Hariri, Mobasher, and Burke]{hariri2012context}
Negar Hariri, Bamshad Mobasher, and Robin Burke.
\newblock Context-aware music recommendation based on latenttopic sequential
  patterns.
\newblock In \emph{Proceedings of the sixth ACM conference on Recommender
  systems}, pages 131--138. ACM, 2012.

\bibitem[Hirai et~al.(2016)Hirai, Doi, and Morishima]{hirai2016musicmixer}
Tatsunori Hirai, Hironori Doi, and Shigeo Morishima.
\newblock Musicmixer: Automatic dj system considering beat and latent topic
  similarity.
\newblock In \emph{International Conference on Multimedia Modeling}, pages
  698--709. Springer, 2016.

\bibitem[Hirai et~al.(2018)Hirai, Doi, and Morishima]{hirai2018latent}
Tatsunori Hirai, Hironori Doi, and Shigeo Morishima.
\newblock Latent topic similarity for music retrieval and its application to a
  system that supports dj performance.
\newblock \emph{Journal of Information Processing}, 26:\penalty0 276--284,
  2018.

\bibitem[Hu and Saul()]{huprobabilistic}
Diane~J Hu and Lawrence~K Saul.
\newblock A probabilistic topic model for music analysis.

\bibitem[Kim et~al.(2009)Kim, Narayanan, and Sundaram]{kim2009acoustic}
Samuel Kim, Shrikanth Narayanan, and Shiva Sundaram.
\newblock Acoustic topic model for audio information retrieval.
\newblock In \emph{Applications of Signal Processing to Audio and Acoustics,
  2009. WASPAA'09. IEEE Workshop on}, pages 37--40. IEEE, 2009.

\bibitem[Kim et~al.(2012)Kim, Georgiou, and Narayanan]{kim2012latent}
Samuel Kim, Panayiotis Georgiou, and Shrikanth Narayanan.
\newblock Latent acoustic topic models for unstructured audio classification.
\newblock \emph{APSIPA Transactions on Signal and Information Processing}, 1,
  2012.

\bibitem[McFee et~al.(2015)McFee, Raffel, Liang, Ellis, McVicar, Battenberg,
  and Nieto]{mcfee2015librosa}
Brian McFee, Colin Raffel, Dawen Liang, Daniel~PW Ellis, Matt McVicar, Eric
  Battenberg, and Oriol Nieto.
\newblock librosa: Audio and music signal analysis in python.
\newblock In \emph{Proceedings of the 14th python in science conference}, pages
  18--25, 2015.

\bibitem[Shalit et~al.(2013)Shalit, Weinshall, and Chechik]{shalit2013modeling}
Uri Shalit, Daphna Weinshall, and Gal Chechik.
\newblock Modeling musical influence with topic models.
\newblock In \emph{International Conference on Machine Learning}, pages
  244--252, 2013.

\bibitem[Tzanetakis and Cook(2002)]{tzanetakis2002musical}
George Tzanetakis and Perry Cook.
\newblock Musical genre classification of audio signals.
\newblock \emph{IEEE Transactions on speech and audio processing}, 10\penalty0
  (5):\penalty0 293--302, 2002.

\bibitem[Wells(1992)]{wells1992origins}
Paul~F Wells.
\newblock Origins of the popular style: The antecedents of twentieth-century
  popular music, 1992.

\end{thebibliography}
\bibliographystyle{plainnat}

\end{document}